\documentclass[10pt,conference,letterpaper]{IEEEtran}
\IEEEoverridecommandlockouts

\usepackage[T1]{fontenc}
\usepackage[utf8]{inputenc}
\usepackage{amsmath,amssymb}
\usepackage{booktabs}
\usepackage{cite}
\usepackage{graphicx}
\usepackage{siunitx}
\usepackage{tabularx}
\usepackage{tikz}
\usetikzlibrary{arrows.meta,positioning}
\usepackage{xurl}
\usepackage[hidelinks,bookmarks=false]{hyperref}

\title{Channel Knowledge Maps for FR3 UAV Base Stations: Dataset and ML-based Modeling}

\author{\IEEEauthorblockN{Guillem Moreno Garcia, Sergi Abadal, Evgenii Vinogradov}
\IEEEauthorblockA{Universitat Polit\`ecnica de Catalunya (UPC), Barcelona, Spain\\
guillem.moreno.garcia@estudiantat.upc.edu}
\thanks{This work was supported by the Ramón y Cajal grant RYC2024-051003-1 and the I+D+i project BLOSSOMS (grant PID2024-158530OB-I00) funded by MICIU/AEI/10.13039/501100011033 and EU.}}

\begin{document}
\maketitle

\begin{abstract}
This paper introduces a dense upper mid-band Frequency Range (FR3) Channel Knowledge Map (CKM) dataset for Uncrewed Aerial Vehicles (UAVs) with an on-board base station (UxNB). It contains 16\,180 CKM samples from 3\,236 urban locations in 66 cities at {7.125}\,{GHz} and UxNB heights of up to 500 meters. Every 513$\times$513 sample stores the building geometry, the line-of-sight (LOS) state, attenuation, delay spread, angular spread, and variable transmitter height. As a focused first use-case, we propose an interpretable height-aware attenuation model. Its LOS branch combines free-space loss, two-ray term, and compact data-based height/range calibration. Its non-LOS (NLOS) branch combines a COST\,231-Hata basis with local environment-specific linear calibration. The proposed model reaches {1.93}\,{dB} overall root mean square error, with {1.74}\,{dB} in LOS and {3.54}\,{dB} in NLOS on over 500 unique locations in 14 unseen cities (2\,590 samples in total). 
The result provides attenuation data for FR3 UxNB deployments and a physical anchor for future FR3 modeling efforts.
\end{abstract}


\section{Introduction}
Uncrewed Aerial Vehicles (UAVs) with an on-board base station (UxNB) are often seen as one of the key enablers of future 6G networks due to their mobility and better Line-of-Sight (LOS) channels~\cite{Vinogradov2018,Geraci2022ntn}. However, UxNB deployment faces challenges such as location optimization and real-time channel state estimation on resource-constrained UxNBs. Environment-aware communication enabled by Channel Knowledge Maps (CKMs) aims to resolve these challenges~\cite{ckmtutorial2024}. A CKM is a site-specific database containing the locations of transmitters and/or receivers, providing location-specific channel knowledge useful to enhance environment-awareness and facilitate channel acquisition or obviate the need for it entirely. Any relevant information can be stored in a CKM, including but not limited to channel gains, multipath delays/angles, or even the full channel impulse response~\cite{zengx2021ckm}.

CKMs can be obtained via measurements or channel modeling. Measurements provide the most accurate results, but may be cumbersome. The models used for simulating wireless channels range from a fully stochastic to fine-grained deterministic Ray Tracing (RT) model depending on the requirements for precision and the geographical scale. Most available CKM datasets~\cite{dataset2212,ckmimagenet2025} rely on RT simulated data. Since RT simulators are known to be resource-demanding, faster learning-based simulation approaches have become popular \cite{Chen2017rem,radiounet2020,pmnet2023, pmnet_icassp2023}. Prediction of dense channel gain maps is commonly approached as an image-to-image regression problem: RadioUNet~\cite{radiounet2020} and PMNet~\cite{pmnet2023, pmnet_icassp2023} systems demonstrate the strength of convolutional encoder--decoder models when city topology and sparse or simulated channels are available. 

Recently, the upper mid-band Frequency Range (FR3, between 7 and 24\,GHz) has started to gain attention as a promising spectrum candidate to enhance both coverage and capacity of 6G networks~\cite{Cui2024}. FR3 offers a balance between the reliable propagation of sub-6\,GHz bands (FR1) and the large bandwidth of millimeter-wave bands (FR2). FR3 is very attractive for UxNB deployments that have better LOS conditions but still must reliably function in Non-LOS (NLOS) scenarios.  To assess and harness the potential of UxNB-enabled networks in FR3, we need accurate channel data in diverse urban environments. Unfortunately, to our knowledge, there are no publicly available UxNB CKM datasets in FR3.

\begin{table}[b]
\vspace{-15pt}
\centering
\caption{Comparison of Different CKM Datasets}
\vspace{-2pt}
\begin{tabular}{@{}c|c|c|c|c@{}}
\toprule
Properties  & Ours   &  \cite{dataset2212}  &  \cite{ckmimagenet2025} &\cite{radiodiff2025}   \\ \hline 
Frequency Range & FR3   & FR1 & FR2 & FR1\\ \hline 
Elevated Tx height & yes   & no & no &no\\ \hline 
Number of Layouts & 3\,236  & 701  & 40\,000 & 701 \\ \hline 
Size of Sample  & 513$\times$513& 256$\times$256  & 128$\times$128 &256$\times$256 \\  \hline 
Horizontal Resolution & 1 m & 1 m  & 2 m  & 1 m\\ \hline 
Real Buildings Footprint  & yes   & yes     & yes   &yes \\  \hline 
Real Buildings Height & yes & no & no     & no  \\ 
\bottomrule
\end{tabular}
\label{tab:datasets}
\end{table}

\paragraph{Limitations of the existing CKM datasets}
The existing CKM datasets are summarized in Table~\ref{tab:datasets}. It can be seen that the datasets i) contain only FR1 or FR2 channels; ii) do not use UxNB-specific geometries (coverage size and transmitter (Tx) heights); iii) use non-accurate 3D city models. 

\begin{figure*}
\centering
\includegraphics[width=0.85\textwidth]{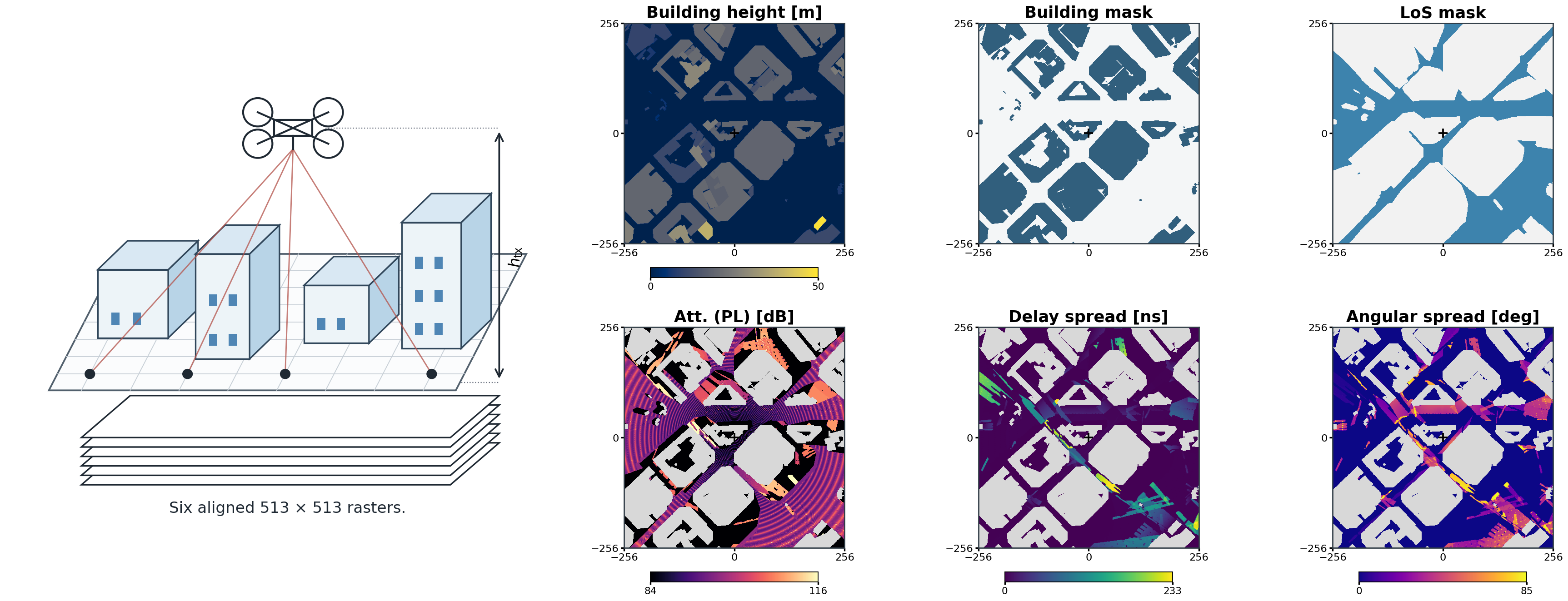}
\caption{An example of a CKM sample from Barcelona. Left: Conceptual Air-to-Ground propagation geometry from an elevated UxNB across a 3D urban topology. Right: The corresponding six aligned $513 \times 513$ layers comprising $L=2$ environmental and $F=4$ channel layers.}
\label{fig:ckm_sample}
\vspace{-8pt}
\end{figure*}

CKM for UxNB should contain information on a typical cell coverage diameter (300--500\,m) with a spatial resolution that is capable of capturing important channel variations (20--80 wavelengths) of channel parameters~\cite{saboor26}. Since UxNBs can use heights up to 500\,m, the resulting channel is very different from terrestrial channels~\cite{khawaja2019survey,Vinogradov2018}. The only existing dataset containing some sort of 3D channel is UrbanRadio3D~\cite{radiodiff2025} using receiver heights of up to 20\,m but it is not sufficient. Moreover, Air-to-Ground (A2G) channel modeling requires accurate information about building height~\cite{vinogradov2026upsim}. The existing datasets use OpenStreetMap (OSM) city layouts. Although OSM contains accurate information about building footprints, critical building height information is only available in less than 10\% of cases~\cite{Bernard2022}.

\paragraph{Limitations of the existing CKM modeling approaches}
Existing learning-based channel map estimation frameworks suffer from several architectural and practical limitations. First, dense attenuation-map prediction is commonly approached via pure image-to-image regression, employing heavy, over-parameterized convolutional networks to rediscover propagation physics (e.g., free-space path loss) independently for every environment. Second, these architectures typically assume fixed terrestrial geometries and fail to reproduce the effect of varying transmitter altitudes typical for UxNB deployments. Finally, pure deep learning approaches struggle with spatial generalization; they often require local fine-tuning when deployed in unseen cities and urban topologies. 

\paragraph{Beyond state-of-the-art contributions}
In this paper, we address these gaps. The main contribution thus is two-fold:
\begin{enumerate}
    \item We introduce an FR3 (7.125 GHz) Channel Knowledge Map dataset\footnote{\url{https://doi.org/10.5281/zenodo.21839567}} generated via ray tracing simulations in 3\,236 city layouts across 66 diverse cities. This dataset relies on complete, global-scale 3D building data~\cite{Che2024globfp,Zhu2025atlas}. UxNB applications are enabled by using Tx heights of up to 500 meters (5 random heights per layout resulting in a total of 16\,180 CKM samples). 
    \item We propose a height-aware channel prediction framework\footnote{\url{https://github.com/unworthyzeus/CKMAttenuationPriors/tree/main}} that explicitly separates LOS and NLOS propagation using local city geometry. By calibrating deterministic baseline models (e.g., free-space, two-ray, COST\,231-Hata) via lightweight regression, the proposed approach achieves highly accurate, zero-shot spatial generalization to completely unseen cities while offering computational speedups.
\end{enumerate}

The paper is organized into dataset description, modeling, evaluation, and conclusions (Sections~\ref{sec:ckm_dataset}--\ref{sec:conclusions}).

\textbf{Notation:} Boldface uppercase letters $\mathbf{X}$ denote matrices or tensors, while ${X}_{p}$ represents the scalar entry of matrix $\mathbf{X}$ at position $p = (i, j)$. Boldface lowercase letters $\mathbf{x}$ denote vectors, and italic letters $x$ denote scalar variables. The operator $\odot$ represents the Hadamard (element-wise) matrix product, and $\mathbf{x}^\top$ denotes the transpose of vector $\mathbf{x}$.

\section{CKM: Definition and Dataset Description}
\label{sec:ckm_dataset}

An example of a CKM sample from our dataset is illustrated in Fig.~\ref{fig:ckm_sample}. In the following, we will use the terms UxNB and transmitter interchangeably.

\subsection{CKM Definition and Mapping Formulation}
\label{sec:ckm_def}
We define a CKM sample as a spatial representation combining environmental features, channel parameters, and operational metadata. For any given CKM sample $n \in \{1, \dots, N\}$, the propagation environment is represented as a tensor $\mathbf{E}_n \in \mathbf{E} \subset \mathbb{R}^{H \times W \times L}$, where $H$ and $W$ denote the spatial grid dimensions (spanning element indices $1$ to $513$), and $L$ is the number of environmental layers (e.g., building heights, terrain information). The radio propagation characteristics are represented as a channel state tensor $\mathbf{C}_n \in \mathbf{C} \subset \mathbb{R}^{H \times W \times F}$, storing $F$ channel parameters at the receiver grid (e.g., attenuation, multipath delay spread, angular parameters). Operational metadata (e.g., carrier frequencies, UxNB location information) are defined via a metadata vector $\mathbf{m}_n \in \mathcal{M} \subset \mathbb{R}^K$.

Hence, CKM estimation can be framed as finding a parametric mapping function $\mathbf{\Phi}_{\boldsymbol{\theta}}$:
\begin{equation}
\mathbf{\Phi}_{\boldsymbol{\theta}} : \mathbf{E} \times \mathcal{M} \to \mathbf{C},
\label{eq:mapping_generic}
\end{equation}
where parameters $\boldsymbol{\theta}$ are optimized such that the estimated channel tensor $\widehat{\mathbf{C}}_n = \mathbf{\Phi}_{\boldsymbol{\theta}}(\mathbf{E}_n, \mathbf{m}_n)$ approximates the ground-truth target $\mathbf{C}_n$.

\subsection{The FR3 UxNB CKM Dataset}
\label{sec:dataset_details}
We build a comprehensive benchmark dataset using high-precision deterministic RT simulations.

\subsubsection{3D urban environment modeling ($\mathbf{E}$)}
\label{sec:env_model}
The environmental tensor $\mathbf{E}_n$ in our dataset relies on global-scale 3D building datasets~\cite{Che2024globfp,Zhu2025atlas}. For UxNB deployments, this offers a clear advantage over relying on OSM data popular in literature. OSM is highly valuable for footprint extraction, but height-related tags such as {"height"} and {"building:levels"} are optional and their availability is less than 10\%~\cite{Bernard2022}. By contrast, combining 3D-GloBFP~\cite{Che2024globfp} and GlobalBuildingAtlas~\cite{Zhu2025atlas} provides consistent building height information\footnote{Both datasets claim global coverage but some geographical locations are missing. For example, GlobalBuildingAtlas~\cite{Zhu2025atlas} does not contain the map of Barcelona and some other cities in the EU, while it covers many cities in East Asia missing in 3D-GloBFP~\cite{Che2024globfp}. In this paper, we first searched locations in the more recent GlobalBuildingAtlas and used 3D-GloBFP as the backup.}. This is crucial for UxNBs that depend heavily on deployment height~\cite{Vinogradov2018}.

We extract city layouts of 3\,236 urban locations in 66 diverse cities worldwide to cover a range of propagation topologies (from open low-rise suburbs to dense high-rise downtown areas). Each rasterized environmental map $\mathbf{E}_n$ spans a spatial grid of $H = W = 513$ at a horizontal resolution of $1\text{~m}$. The $L=2$ environmental channels encode:
\begin{enumerate}
    \item Building height profile maps $\mathbf{H}$ indicating building height in meters ($0\text{~m}$ over ground level).
    \item Binary building occupancy masks $\mathbf{B}$ indicating presence of structural obstacles.
\end{enumerate}

\subsubsection{Ray Tracing channel simulation ($\mathbf{C}$)}
\label{sec:rt_sim}
High-precision ray tracing simulations are conducted to generate the spatially consistent channel tensors $\mathbf{C}_n$. To avoid using separate software for 3D model construction (e.g., Blender) and for RT (e.g., Wireless InSite or Sionna), SiteViewer and RT capabilities of MATLAB are used in this work. 

RT simulation parameters are summarized in Table~\ref{tab:rt_params}. The $7.125\text{~GHz}$ frequency was selected since the recent World Radiocommunication Conference (WRC-23) indicated it as the only FR3 test band available in all ITU regions. The $1\text{-m}$ resolution corresponds to 24 wavelengths, ensuring spatial coherence~\cite{saboor26}. The other parameters are selected to be transferable to typical UxNB deployments. The total size of the simulated area is larger than in other datasets~\cite{dataset2212,ckmimagenet2025} and corresponds to an increased UxNB coverage footprint. Moreover, the UxNB is always located in the middle of the layout at elevated heights (10~m to 500~m, sampling 5 heights per location) to reflect the focus on UxNB-centric scenarios. This results in a total dataset size of $N =$ 16\,180 samples ($3\,236 \text{ layouts} \times 5 \text{ Tx heights}$). The receivers are placed at ground level in all non-building coordinates. We use the number of ray interactions from CKM literature~\cite{dataset2212}. The resulting channel state tensors $\mathbf{C}_n$ store radio parameters across $F=4$ channels:
\begin{enumerate}
    \item Binary visibility mask $\mathbf{V}$: LOS ($1$) vs. NLOS ($0$) states.
    \item Channel attenuation in dB.
    \item Multipath Root Mean Square (RMS) delay spread in ns.
    \item  Multipath RMS angular spread in degrees.
\end{enumerate}
In-building elements are assigned a value of $0$ in $\mathbf{C}_n$ and should be masked out during training and evaluation.

\subsubsection{Metadata ($\mathbf{m}$)}
\label{sec:metadata}

Because carrier frequency, receiver height, and horizontal transmitter location (fixed at the grid center) are constant across all simulations, the metadata vector $\mathbf{m}_n$ consists of:
\begin{enumerate}
    \item Transmitter height $h_{\text{tx}}$ ranging from $10\text{~m}$ to $500\text{~m}$.
    \item Classifying tag mapping the city layout into the closest standard ITU urban topology~\cite{iturp1410} based on building coverage $\alpha$, the density $\beta$ (buildings/km$^2$) as well as the mean building height $\gamma$\,{(m)}:
\end{enumerate}

    \begin{center}
    \vspace{-0.1cm}
    \small
    \begin{tabular}{|l|c|c|c|c|}
    \hline
    {Topology} & {Samples} & $\alpha$ & $\beta$ {} & $\gamma$ \\ \hline
    Suburban    & 7\,435 & 0.1 & 750 & 8  \\ \hline
    Urban       & 7\,350 & 0.3 & 500 & 15 \\ \hline
    Dense Urban & 1\,395 & 0.5 & 300 & 20 \\ \hline
    \end{tabular}
    \vspace{-0.1cm}
    \end{center}
The practical $\hat\alpha$, $\hat\beta$, and $\hat\gamma$ differ from the tabular values. Hence, we compute per-topology distances $d=\|\log[(\hat\alpha,\hat\beta,\hat\gamma)/(\alpha,\beta,\gamma)]\|_2$ and choose the smallest. Equal-weight
log-ratios compare relative deviations and prevent the scale of $\beta$ from
dominating. 
    
\begin{table}[]
\centering
\caption{Ray Tracing Simulation Parameters}
\label{tab:rt_params}
\begin{tabular}{lc}
\hline
{Parameter} & {Value} \\ \hline
Carrier frequency, $f$ & $7.125\,\text{GHz}$ \\
Antenna model & Isotropic \\
Sample dimensions & $513\,\text{m} \times 513\,\text{m}$\\
Spatial resolution & $1\,\text{m}$ \\
Ground receiver height, $h_{\text{rx}}$ & $1.5\,\text{m}$ \\
UxNB heights, $h_{\text{tx}}$ & $10\,\text{m}$ to $500\,\text{m}$ \\
UxNB heights per layout & 5 \\
Number of layouts & 3\,236 \\
Total CKM samples, $N$ & 16\,180 \\
Number of ray interactions & 2 \\
\hline
\end{tabular}
\vspace{-0.5cm}
\end{table}

    \section{Height-Aware CKM modeling}\label{sec:ckm_model}
    
    \begin{figure}[b]
    \centering
    \vspace{-0.2cm}
    \resizebox{\columnwidth}{!}{%
    \begin{tikzpicture}[
    node distance=5mm and 6mm,
    box/.style={draw,rounded corners,align=center,minimum height=7mm,font=\scriptsize},
    arr/.style={-{Latex[length=2mm]},thick}]
    \node[box,fill=gray!10] (input) {Environment model $\mathbf{E}$ \\ LOS/NLOS mask $\mathbf{V}$\\ Metadata $\mathbf{m}$};
    \node[box,fill=blue!9,right=of input,yshift=5mm] (los) {LOS: free-space and two-ray\\+\\geometry calibration};
    \node[box,fill=orange!12,right=of input,yshift=-10mm] (nlos) {NLOS: COST\,231-Hata\\+\\ geometry and environment \\calibration};
    \node[box,fill=green!10,right=10mm of los,yshift=-5mm] (switch) {LOS/NLOS \\ combining};
    \node[box,fill=purple!9,right=of switch] (out) {Channel \\Attenuation Map};
    \draw[arr] (input) -- (los);
    \draw[arr] (input) -- (nlos);
    \draw[arr] (los) -- (switch);
    \draw[arr] (nlos) -- (switch);
    \draw[arr] (switch) -- (out);
    \end{tikzpicture}}
    \caption{The proposed solution uses separate LOS and NLOS propagation models calibrated with training CKM data.}
    \label{fig:CKM_model}
    \end{figure}
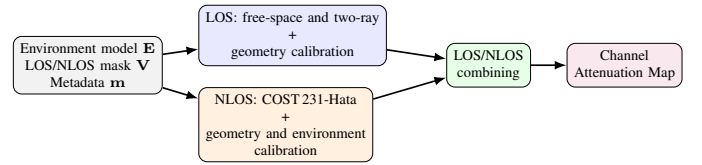
    Channel attenuation modeling requires resolving two propagation regimes: (i) LOS channels experience free-space path loss and two-ray coherent ground reflections, and (ii) NLOS channels are dominated by multipath propagation affected by local buildings (especially near the receiver). Furthermore, as the UxNB altitude $h_{\text{tx}}$ increases, the probability of LOS and the dominant propagation mechanisms change. Standard 2D image-to-image networks struggle to natively capture these altitude-dependent 3D spatial variations. 
    
    To address this, we propose a propagation-informed CKM attenuation simulator that explicitly separates the environment into LOS and NLOS areas (see Fig.~\ref{fig:CKM_model}). The model implements the attenuation-focused mapping function $\mathbf{\Phi}_{\boldsymbol{\theta}}(\mathbf{E}_n, \mathbf{m}_n)$ by combining distinct LOS and NLOS models:
    \begin{equation}
    \Lambda=\mathbf{V} \odot \Lambda_\text{LOS} + (1-\mathbf{V}) \odot \Lambda_\text{NLOS},
    \label{eq:combined}
    \end{equation}
    where $\mathbf{V}$ is the dataset visibility mask. Reported accuracy assumes this mask; its generation via RT or shadow projection~\cite{vinogradov2026shadow,vinogradov2026upsim} is excluded from timing but it's in the same order of magnitude as the proposed solution.
    
    Using CKM training data, we employ geometry-dependent calibration for LOS and NLOS branches. The proposed LOS attenuation model relies on the free-space and two-ray models corrected with data-learned terms. The proposed NLOS model takes into account the ITU topology characteristics and uses a CKM data-corrected version of the COST\,231-Hata model. 
    
    Given the deployment geometry ($h_{\text{tx}}$, $h_{\text{rx}}$, and the fixed horizontal UxNB position), we pre-compute spatial geometry matrices over the $513 \times 513$ grid: horizontal distance $\mathbf{D}_{\text{2D}}$, direct 3D distance $\mathbf{D}_{\text{3D}}$, ground-reflected path length $\mathbf{D}_{\text{GR}}$, and elevation angle $\mathbf{\Theta}_{\text{2D}}$. For any receiver location $p = (i, j)$, the scalar geometric parameters ($d_{\text{2D}}, d_{\text{3D}}, d_{\text{GR}}, \theta$) are indexed directly from these matrices. For notational brevity, we omit the spatial index $p$ in subsequent expressions.
    
    \subsection{Line-of-Sight Branch}
    \label{sec:los}
    For LOS propagation, the attenuation model combines direct Free-Space Path Loss (FSPL) $\Lambda_\text{FSPL}$ with a coherent ground-reflection term $C_{\text{2ray}}$ to capture the two-ray propagation effects. Additionally, the model is calibrated based on CKM data as
    \begin{equation}
    \Lambda_{\text{LOS}} = \Lambda_\text{FSPL}(d_{\text{3D}}) + C_{\text{2ray}} + \zeta(h_{\text{tx}}) + r(d_{\text{2D}},h_{\text{tx}}),
    \label{eq:pl_los}
    \end{equation}
    where $\zeta(h_{\text{tx}})$ is a height-dependent scalar bias correction and $r(d_{\text{2D}},h_{\text{tx}})$ is a residual component capturing the effects introduced by horizontal distance and transmitter height.
    
    The free-space term follows the Friis distance dependence, while the coherent
    correction retains the phase difference between the direct and reflected paths
    \cite{rappaport}:
    \begin{equation} \Lambda_\text{FSPL}=20\log_{10}\Big(\frac{4\pi d_\text{3D} f}{c}\Big),\label{eq:fspl}
    \end{equation}
    \begin{equation}
    C_{\mathrm{2ray}}=-20\log_{10}\left|1+\rho(h_{\mathrm{tx}})
    \frac{d_{\mathrm{3D}}}{d_{\mathrm{GR}}}
    e^{-j[k(d_{\mathrm{GR}}-d_{\mathrm{3D}})+\phi(h_{\mathrm{tx}})]}\right|,
    \label{eq:tworay}
    \end{equation}
    where $k$ is the free-space wavenumber, $\rho$ is an effective reflected-field amplitude, and $\phi$ is a fitted phase offset. The calibration terms are obtained in two steps.
    Firstly, in each \SI{5}{m} height bin, candidates $\hat\rho\in\{0,0.05,\ldots,1.5\}$ and 48 uniformly spaced phases $\hat{\phi}$ in $[-\pi,\pi)$ are evaluated. For a candidate set (i.e., $\hat\zeta,\hat\rho,\hat\phi$), the remaining bias is calculated as
    \begin{equation}
    \hat\zeta(\hat\rho,\hat\phi)=\frac{1}{|B_\text{LOS}|}\sum_{i\in B_\text{LOS}}
    \left[y_i-\Lambda_{\text{FSPL},i}-C_{\mathrm{2ray},i}(\hat\rho,\hat\phi)\right],
    \label{eq:los-fit}
    \end{equation}
    where $B_\text{LOS}$ is the set of valid LOS receivers and $y$ is the ground truth attenuation. The combination of $\hat\zeta,\hat\rho,\hat\phi$ minimizing the bin Root Mean Square Error (RMSE) is retained.
    Secondly, the remaining radial calibration term $r(d_{\text{2D}},h_\text{tx})$ is calculated as the mean of the residual error.

    In implementation terms, the LOS calibration is a lookup table that contains height- and distance-dependent $\rho$, $\phi$, $\zeta$, and $r$. Inference linearly interpolates neighboring entries.
    
    \subsection{Non-Line-of-Sight Branch}
    \label{sec:nlos}
    
    For NLOS receiver locations, our modeling approach relies on an empirical COST\,231-Hata baseline $\Lambda_C$, augmented by local topographical features derived from the environment map $\mathbf{E}$. The predicted NLOS attenuation is given by:
    \begin{equation}
    \Lambda_{\text{NLOS}} = \mathbf{x}^\top \hat{\mathbf{w}}_q,
    \label{eq:pl_nlos}
    \end{equation}
    where $\mathbf{x} \in \mathbb{R}^{14}$ is a spatial context feature vector and $\hat{\mathbf{w}}_q$ is the height- and topology-dependent calibrated weight vector.
    
    Evaluated at $f_{\text{MHz}} = 7\,125$\footnote{Since the COST\,231-Hata formulation was designed for sub-3~GHz terrestrial links, we use it as a regression feature with coefficients fitted on training data.}, the term $\Lambda_C$ is computed as:
    \begin{align}
    \Lambda_C &= 46.3 + 33.9 \log_{10} f_{\text{MHz}} - 13.82 \log_{10} h_{\text{tx}} - a(h_{\text{rx}}) \nonumber \\
    &\qquad\quad + \eta(h_{\text{tx}}) \log_{10} (d_{\text{3D}}/1000) + 3,\\ a(h_{\text{rx}}) &= (1.1 \log_{10} f_{\text{MHz}} - 0.7) h_{\text{rx}} - (1.56 \log_{10} f_{\text{MHz}} - 0.8), 
    \label{eq:cost}
    \end{align}
    \begin{equation}
        \eta(h_{\text{tx}}) = 44.9 - 6.55 \log_{10} h_{\text{tx}}.
    \end{equation}
    To account for the local environment, we derive 13 contextual features from the building height grid $\mathbf{H}$ and building presence map $\mathbf{B}$ in $\mathbf{E}$. Let $\mathbf{G} = 1 - \mathbf{B}$ be valid outdoor receiver locations and $\mathbf{N} = (1 - \mathbf{V}) \odot \mathbf{G}$ NLOS-ground locations. We use two scales for topographical features ($\pm$7 and 20\,m), the resulting window size $l \in \{15, 41\}$. The reflect-padded box mean operator is defined as $\mathcal{A}_l[z] = l^{-2} \sum_{q \in \mathcal{W}_l} z_q$.
    
    The full feature vector $\mathbf{x}$ is constructed as:
    \begin{equation}
    \begin{aligned}
    \mathbf x=[&\Lambda_{C},\ell_d,\delta_{15},\delta_{41},h_{15},h_{41},
    \delta_{41}\ell_d,\\
    &n_{15},n_{41},n_{41}\ell_d,\sigma_{\mathrm{sh}},\theta',
    n_{41}\theta',1]^{\top}.
    \end{aligned}
    \label{eq:nlos-vector}
    \end{equation}
    where the components represent:
    \begin{itemize}
        \item Baseline and Distance: The COST\,231-Hata baseline $\Lambda_C$ and log-radial decay $\ell_d = \ln(1 + d_{\text{2D}})$.
        \item Structural Context: Building density $\delta_l = \mathcal{A}_l[\mathbf{B}]$, normalized building heights $h_l = \mathcal{A}_l[\mathbf{H} \cdot \mathbf{B}] / 90$, and shadowed ground fraction $n_l = \mathcal{A}_l[\mathbf{N}]$ evaluated within local ($l=15$) and regional ($l=41$) windows.
        \item Elevation effects: Normalized path elevation angle $\theta' = \operatorname{clip}(\theta / 90, 0, 1)$ and a deterministic shadowing proxy $\sigma_{\text{sh}} = 2.3197 \left[\max(90 - \theta, 0)\right]^{0.2361}$ as in~\cite{feng2006a2g}.
        \item Interaction Cross-Terms: Density--range $\delta_{41}\ell_d$, blockage--range $n_{41}\ell_d$, and blockage--elevation $n_{41}\theta'$ interaction terms, alongside the scalar bias unit $1$.
    \end{itemize}
    
    The regime weight vectors $\hat{\mathbf{w}}_q$ are calibrated offline using CKM training samples in two steps. Firstly, the scenes are partitioned into $q$ regimes resulting from combinations of the three ITU topologies and three Tx altitude $h_{\text{tx}}$ (nine in total). Altitude regimes are binned into \textit{low} ($h_{\text{tx}} \le 60\text{~m}$), \textit{medium} ($60\text{~m} < h_{\text{tx}} \le 100\text{~m}$), and \textit{high} ($h_{\text{tx}} > 100\text{~m}$). Secondly, for each regime $q$, at most 1\,024 NLoS channel realizations per training sample are fed to the ridge model. For training stability during offline calibration, let $\mathbf{z}_i$ denote the standardized counterpart of the spatial context feature vector $\mathbf{x}_i$ for sample $i$:
        \begin{equation}
        \hat{\mathbf{w}}_q = \arg\min_{\mathbf{w}} \sum_{i \in q} \left( \mathbf{z}_i^\top \mathbf{w} - y_i \right)^2 + 10^{-2} \|\mathbf{w}_{\neg b}\|_2^2,
        \label{eq:nlos-ridge}
        \end{equation}
        where $\mathbf z_i$ contains standardized non-intercept features, and the $L_2$ term penalizes only $\mathbf w_{\neg b}$ while omitting the bias (intercept). 

\section{Experiment}\label{sec:results}
\begin{figure}[!t]
\centering
\includegraphics[width=0.8\columnwidth]{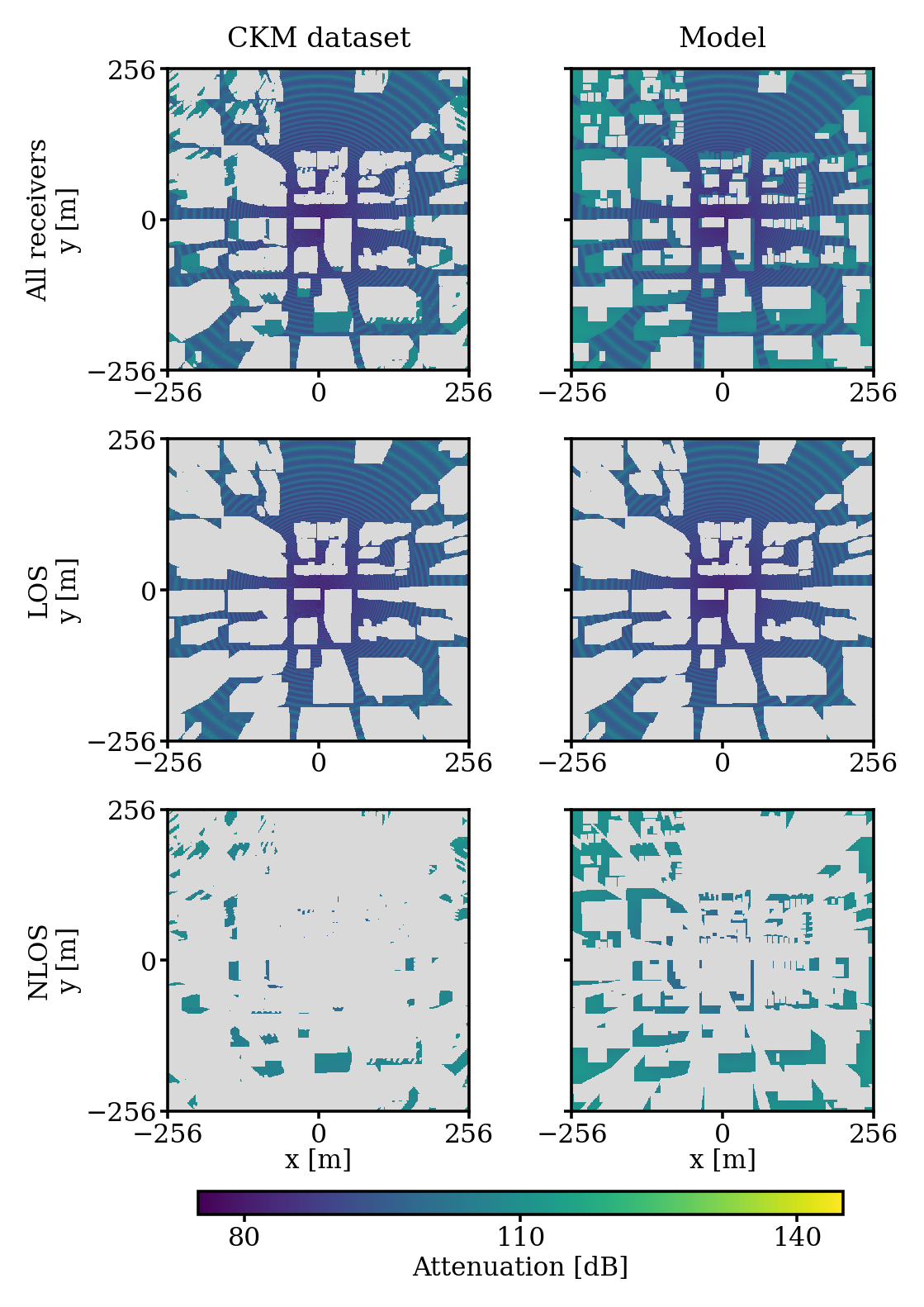}
\caption{An example from Vancouver
(``sample\_15262'', $h_{\mathrm{tx}}=${50.30}\,{m}, Dense urban topology, low-transmitter regime). 
The model reproduces attenuation over the unseen layout with {2.18}\,{dB} RMSE. The sample contains 119\,180 LOS and 25\,607 NLOS channels, modeled with {2.10}\,{dB} and {2.52}\,{dB} RMSE. }
\label{fig:output_example}
\vspace{-0.3cm}
\end{figure}

\subsection{Experimental Setup and Evaluation Protocol}
The dataset split contains 10\,840 samples from 37 training cities, 2\,750 samples from 15 validation cities, and 2\,590 samples from 14 final test cities. No city is moved between these partitions to ensure strict spatial generalization. Training procedures described in Section~\ref{sec:ckm_model} use only training cities, where all nine topology and height regimes are fitted on the 10\,840 CKM training samples. The validation subset is reserved for development checks and component ablation. Final performance results use the unseen 14-city test subset.

After excluding building interiors and outdoor channels lacking valid ray tracing outputs, the test dataset evaluates over $4.23 \times 10^8$ ground receivers. Of these, only $7.4\%$ are in NLOS states due to the elevated UxNB geometry. Although the element-weighted overall score is therefore LOS-dominated, 424 of the 2\,590 test maps contain more than $25\%$ NLOS receivers, motivating our separate regime-specific evaluations. An example of the model output can be seen in Fig.~\ref{fig:output_example}.

\subsection{Prediction Accuracy and Height Generalization}
The prediction performance on the unseen test cities across different ITU topologies is summarized in Table~\ref{tab:itu-topology}. Overall, the proposed simulator achieves a global RMSE of \SI{1.93}{dB} (\SI{1.74}{dB} for LOS and \SI{3.54}{dB} for NLOS). While Suburban topology exhibits the lowest total error of \SI{1.70}{dB}, Dense Urban layouts present a harsher NLOS environment with a higher error of \SI{4.13}{dB}. Per-city RMSE ranges from \SI{1.61}{dB} in Halifax to \SI{2.59}{dB} in Osaka, with an unweighted 14-city mean of \SI{1.98}{dB}. The fitted coefficient vectors also change substantially across the ITU topologies, so one global set of slopes would mix distinct relationships. These observations further validate the need for regime calibrations by topology.

Table~\ref{tab:itu-topology} also breaks down the final test error by UxNB altitude regimes. As altitude increases from \SI{10}{m} to \SI{500}{m}, the overall RMSE decreases monotonically from \SI{2.14}{dB} to \SI{1.71}{dB}. This aligns with the observations that higher-altitude channels are usually more stable and predictable~\cite{Vinogradov2018,khawaja2019survey}.

\begin{table}
\centering
\caption{RMSE (in decibels) by ITU topology}
\label{tab:itu-topology}
\scriptsize
\begin{tabular}{lcccc}
\toprule
Topology  & Overall & LOS & NLOS & Test Samples \\
\midrule
Suburban & 1.70 & 1.60 & 3.16 & 1\,455\\
Urban & 2.30 & 2.00 & 3.65 & 940\\
Dense urban & 2.57 & 2.11 & 4.13 & 195\\
\midrule
\midrule
UxNB height & Overall & LOS & NLOS & Test samples \\
\midrule
Low ($h_{\mathrm{tx}}\le\SI{60}{m}$) & 2.14 & 1.79 & 3.75 & 899 \\
Medium ($\SI{60}{m}<h_{\mathrm{tx}}\le\SI{100}{m}$) & 1.97 & 1.77 & 3.51 & 773 \\
High ($h_{\mathrm{tx}}>\SI{100}{m}$) & 1.71 & 1.67 & 2.76 & 918 \\
\midrule
All topologies and heights & 1.93 & 1.74 & 3.54 & 2\,590 \\
\bottomrule
\vspace{-0.6cm}
\end{tabular}
\end{table}

\begin{table}[t]
\centering
\caption{Comparison with Data-Driven Baseline}
\label{tab:accuracy-context}
\scriptsize
\setlength{\tabcolsep}{4pt}
\begin{tabular}{llccc}
\toprule
Method & Type/Grid & Overall & LOS & NLOS \\
\midrule
{Proposed} & Regime calibrated / $513^2$ & {1.93} & {1.74} & 3.54 \\
GMM U-Net & Neural Baseline / $513^2$  & 4.36 & 4.43 & {3.34} \\
\bottomrule
\vspace{-0.6cm}
\end{tabular}
\end{table}

To assess the performance of pure end-to-end deep learning on this benchmark, we evaluate GMM U-Net with a height-conditioned Gaussian mixture estimator and U-Net decoder, trained with the same training split and evaluated on the same 2\,590 test maps and mask. As shown in Table~\ref{tab:accuracy-context}, the proposed model achieves significantly lower global (\SI{1.93}{dB} vs. \SI{4.36}{dB}) and LOS errors (\SI{1.74}{dB} vs. \SI{4.43}{dB}), demonstrating that our LOS model provides superior stability over the purely data-driven approach. In contrast, the lower NLOS error of the GMM U-Net (\SI{3.34}{dB}) indicates that neural output layers are well-suited to model multipath NLOS channels.

Published CKM and radio map models are designed under fundamentally different conditions (frequencies, heights, generated outputs) and cannot be directly ranked against our $513 \times 513$ aerial dataset. For scale context, terrestrial model RadioUNet~\cite{radiounet2020} reports RMSEs of \SI{1.60}{dB} on fixed-height $256 \times 256$ RadioMapSeer grids; diffusion-based models such as TSRDiff~\cite{tsrdiff2025} achieve \SI{3.62}{dB} under sparse target sampling; and Gaussian process approaches like RO-HMTGP~\cite{rohmtgp2025} report \SI{4.19}{dB}--\SI{6.00}{dB} on empirical RSRP data. The \SI{1.93}{dB} error of our proposed model shows strong prediction accuracy for larger areas with variable UxNB transmitter altitudes.

\subsection{Ablation and Computational Complexity Studies}
Component ablation on the 2\,750-sample validation set confirms the necessity of each calibrated term. Incorporating the coherent ground-reflection term reduces LOS RMSE from \SI{3.79}{dB} to \SI{1.7412}{dB}. For NLOS propagation, using the raw uncalibrated COST\,231-Hata model yields an unacceptable RMSE of \SI{25}{dB} at 7.125~GHz, whereas COST\,231-Hata augmented with the fitted regression coefficients and topology- and height-specific calibration reaches \SI{3.20}{dB} on validation.

Recalibrated trial-and-error tests revealed that a compact model retaining eight of the 14 features (only $l=$41 and no interaction cross-terms) preserves over 90\% of the
complete model's improvement over COST\,231-Hata model.

Table~\ref{tab:compute} reports same-CPU medians on an AMD Ryzen 5 5600X with GPU disabled. RadioUNet/PMNet: 10 untimed and 50 timed inference runs; our model: 100 samples, supplied masks, no file I/O.

\begin{table}[!t]
\centering
\caption{Computational context; reported CPU rows use the same CPU.}
\label{tab:compute}
\scriptsize
\setlength{\tabcolsep}{2.2pt}
\begin{tabular}{@{}lcccc@{}}
\toprule
Method & Grid & Stored scalars & Counted MACs/map & CPU [s] \\
\midrule
Proposed & $513\times513$ & 66.8k+126 & 3.684M$^\ast$ & 0.0602 \\
RadioUNet~\cite{radiounet2020} & $256\times256$ & 13.27M & 12.73G & 0.1339 \\
PMNet~\cite{pmnet2023} & $256\times256$ & 33.34M & 52.71G & 0.4042 \\
SSL-Radio~\cite{sslradio2025} & $256\times256$ & $\approx$15.4--20.4M & $\approx$16.9--21.9G$^\dagger$ & -- \\
MATLAB RT & $513\times513$ & -- & -- & 102.289 \\
\bottomrule
\end{tabular}
\parbox{\columnwidth}{\scriptsize State: 66.8k LOS lookups + 126 NLOS coefficients. $^\ast$Full-grid NLOS dot product only; feature/LOS operations are excluded from MACs but included in time. $^\dagger$SSL-Radio estimate from published layers. RT is a five-layout median with visibility and three channel outputs; timings are orientative and not directly comparable.}
\vspace{-0.6cm}
\end{table}

\section{Conclusions}\label{sec:conclusions}
This paper introduced a dense FR3 (7.125 GHz) UAV CKM dataset providing path loss, delay spread, angular spread, and geometric visibility across 16\,180 simulated samples from 66 global cities. Using this dataset, we proposed a computationally efficient, height-aware channel attenuation prediction framework. By relying on conventional models calibrated on the dataset, the model generates accurate attenuation maps and can generalize to unseen city layouts. The proposed simulator achieves an overall test RMSE of \SI{1.93}{dB} with supplied visibility masks and a median CPU runtime of \SI{0.06}{s} per map.

\bibliographystyle{IEEEtran}
\bibliography{references}

@book{rappaport,
  author    = {Theodore S. Rappaport},
  title     = {Wireless Communications: Principles and Practice},
  edition   = {2},
  publisher = {Prentice Hall},
  year      = {2002},
}

@article{zengx2021ckm,
  author  = {Yong Zeng and Xiaoli Xu},
  title   = {Toward Environment-Aware {6G} Communications via Channel Knowledge Map},
  journal = {IEEE Wireless Communications},
  year    = {2021},
  volume  = {28},
  number  = {3},
  pages   = {84--91},
  doi     = {10.1109/MWC.001.2000327},
}

@article{ckmtutorial2024,
  author  = {Yong Zeng and Junting Chen and Jie Xu and Di Wu and Xiaoli Xu and Shi Jin and Xiqi Gao and David Gesbert and Shuguang Cui and Rui Zhang},
  title   = {A Tutorial on Environment-Aware Communications via Channel Knowledge Map for {6G}},
  journal = {IEEE Communications Surveys \& Tutorials},
  year    = {2024},
  volume  = {26},
  number  = {3},
  pages   = {1478--1519},
  doi     = {10.1109/COMST.2024.3364508},
}

@inproceedings{vinogradov2026shadow,
  author    = {Evgenii Vinogradov and Aymen Fakhreddine and Abdul Saboor and Sergi Abadal and Sofie Pollin},
  title     = {Spatially Consistent Air-to-Ground Channel Modeling and Simulation via {3D} Shadow Projections},
  booktitle = {2026 International Conference on Computing, Networking and Communications},
  year      = {2026},
  doi       = {10.1109/ICNC68183.2026.11416865},
}

@misc{dataset2212,
  author       = {\c{C}. Yapar and R. Levie and G. Kutyniok and G. Caire},
  title        = {Dataset of Pathloss and {ToA} Radio Maps With Localization Application},
  year         = {2022},
  howpublished = {{IEEE} DataPort dataset},
  doi          = {10.21227/0GTX-6V30},
  url          = {https://doi.org/10.21227/0GTX-6V30},
}

@article{ckmimagenet2025,
  author  = {Zijian Wu and Di Wu and Shen Fu and Yuelong Qiu and Yong Zeng},
  title   = {{CKMImageNet}: A Dataset for {AI}-Based Channel Knowledge Map Toward Environment-Aware Communication and Sensing},
  journal = {IEEE Transactions on Communications},
  year    = {2025},
  volume  = {73},
  number  = {12},
  pages   = {14430--14443},
  doi     = {10.1109/TCOMM.2025.3615778}
}

@inproceedings{Cui2024,
  author    = {Zhuangzhuang Cui and Peize Zhang and Sofie Pollin},
  title     = {{6G} Wireless Communications in 7--24 {GHz} Band: Opportunities, Techniques, and Challenges},
  booktitle = {{IEEE} International Symposium on Dynamic Spectrum Access Networks},
  year      = {2025}
}

@article{Bernard2022,
  author  = {J{\'e}r{\'e}my Bernard and Erwan Bocher and Elisabeth Le Saux Wiederhold and Fran{\c{c}}ois Leconte and Val{\'e}ry Masson},
  title   = {Estimation of Missing Building Height in {OpenStreetMap} Data: A French Case Study Using {GeoClimate} 0.0.1},
  journal = {Geoscientific Model Development},
  year    = {2022},
  volume  = {15},
  pages   = {7505--7532}}

@article{Che2024globfp,
  author  = {Yangzi Che and Xuecao Li and Xiaoping Liu and others},
  title   = {{3D-GloBFP}: The First Global Three-Dimensional Building Footprint Dataset},
  journal = {Earth System Science Data},
  year    = {2024},
  volume  = {16},
  pages   = {5357--5374}}

@article{Zhu2025atlas,
  author  = {Xiao Xiang Zhu and Sining Chen and Fahong Zhang and Yilei Shi and Yuanyuan Wang},
  title   = {{GlobalBuildingAtlas}: An Open Global and Complete Dataset of Building Polygons, Heights and {LoD1} {3D} Models},
  journal = {Earth System Science Data},
  year    = {2025},
  volume  = {17}}

@article{radiounet2020,
  author  = {Ron Levie and \c{C}a\u{g}kan Yapar and Gitta Kutyniok and Giuseppe Caire},
  title   = {{RadioUNet}: Fast Radio Map Estimation With Convolutional Neural Networks},
  journal = {IEEE Transactions on Wireless Communications},
  year    = {2021},
  volume  = {20},
  number  = {6},
  doi     = {10.1109/TWC.2021.3054977}}

@inproceedings{pmnet2023,
  author    = {Ju-Hyung Lee and Omer Gokalp Serbetci and Dheeraj Panneer Selvam and Andreas F. Molisch},
  title     = {{PMNet}: Robust Pathloss Map Prediction via Supervised Learning},
  booktitle = {Proc.\ IEEE Global Communications Conference},
  year      = {2023},
  doi       = {10.1109/GLOBECOM54140.2023.10437562}
}

@inproceedings{pmnet_icassp2023,
  author    = {Ju-Hyung Lee and Joohan Lee and Seon-Ho Lee and Andreas F. Molisch},
  title     = {{PMNet}: Large-Scale Channel Prediction System for {ICASSP} 2023 First Pathloss Radio Map Prediction Challenge},
  booktitle = {Proc.\ IEEE International Conference on Acoustics, Speech and Signal Processing ({ICASSP})},
  year      = {2023}
}

@article{radiodiff2025,
  author  = {Xiucheng Wang and Keda Tao and Nan Cheng and Zhisheng Yin and Zan Li and Yuan Zhang and Xuemin Shen},
  title   = {{RadioDiff}: An Effective Generative Diffusion Model for Sampling-Free Dynamic Radio Map Construction},
  journal = {IEEE Transactions on Cognitive Communications and Networking},
  year    = {2025},
  volume  = {11},
  number  = {2},
  pages   = {738--750}}

@techreport{iturp1410,
  author      = {{International Telecommunication Union}},
  title       = {Propagation Data and Prediction Methods Required for the Design of Terrestrial Broadband Radio Access Systems Operating in a Frequency Range from 3 {GHz} to 60 {GHz}},
  institution = {{ITU-R}},
  type        = {Recommendation},
  number      = {P.1410-6},
  year        = {2023},
  month       = aug,
  url         = {https://www.itu.int/rec/R-REC-P.1410-6-202308-I/en},
}

@inproceedings{feng2006a2g,
  author    = {Qixing Feng and Joe McGeehan and Eustace K. Tameh and Andrew R. Nix},
  title     = {Path Loss Models for Air-to-Ground Radio Channels in Urban Environments},
  booktitle = {2006 IEEE 63rd Vehicular Technology Conference},
  year      = {2006},
  volume    = {6},
  pages     = {2901--2905},
  doi       = {10.1109/VETECS.2006.1683399},
}

@ARTICLE{Geraci2022ntn,
  author={Geraci, Giovanni and Garcia-Rodriguez, Adrian and Azari, M. Mahdi and Lozano, Angel and Mezzavilla, Marco and Chatzinotas, Symeon and Chen, Yun and Rangan, Sundeep and Renzo, Marco Di},
  journal={IEEE Communications Surveys \& Tutorials}, 
  title={What Will the Future of {UAV} Cellular Communications Be? A Flight From {5G} to {6G}}, 
  year={2022},
  volume={24},
  number={3},
  doi={10.1109/COMST.2022.3171135}}

@ARTICLE{Vinogradov2018,
  author={Vinogradov, Evgenii and Sallouha, Hazem and De Bast, Sibren and Azari, Mohammad Mahdi and Pollin, Sofie},
  journal={Journal of Mobile Multimedia}, 
  title={Tutorial on {UAVs}: A Blue Sky View on Wireless Communication}, 
  year={2018},
  volume={14},
  number={4},
  pages={395-468},
  doi={10.13052/jmm1550-4646.1443}}

@INPROCEEDINGS{Chen2017rem,
  author={Chen, Junting and Yatnalli, Uday and Gesbert, David},
  booktitle={2017 IEEE International Conference on Communications (ICC)}, 
  title={Learning radio maps for {UAV}-aided wireless networks: A segmented regression approach}, 
  year={2017},
  volume={},
  number={},
  doi={10.1109/ICC.2017.7997333}}

@ARTICLE{saboor26,
  author={Saboor, Abdul and Cui, Zhuangzhuang and Colpaert, Achiel and Vinogradov, Evgenii and Joseph, Wout and Pollin, Sofie},
  journal={IEEE Transactions on Vehicular Technology}, 
  title={Trajectory-Aware Air-to-Ground Channel Characterization for Low-Altitude {UAVs} Using {MaMIMO} Measurements}, 
  year={2026},
  volume={},
  number={},
  pages={1-16},
  doi={10.1109/TVT.2026.3677116}}

@ARTICLE{khawaja2019survey,
  author={Khawaja, Wahab and Guvenc, Ismail and Matolak, David W. and Fiebig, Uwe-Carsten and Schneckenburger, Nicolas},
  journal={IEEE Communications Surveys \& Tutorials}, 
  title={A Survey of Air-to-Ground Propagation Channel Modeling for Unmanned Aerial Vehicles}, 
  year={2019},
  volume={21},
  number={3},
  pages={2361-2391},
  doi={10.1109/COMST.2019.2915069}}

@misc{vinogradov2026upsim,
      title={{UPSim: UxNB} Propagation Simulator for {3D} Map-Driven {FR3} Deployments}, 
      author={Evgenii Vinogradov},
      year={2026},
      eprint={2605.17378},
      archivePrefix={arXiv},
      primaryClass={eess.SP},
      url={https://arxiv.org/abs/2605.17378}, 
}

@inproceedings{tsrdiff2025,
  author    = {Zhanhong Ye and Fan Wu and Cong Zhang and Yitian Shao and Wenhao Fan and Bihua Tang and Yuanan Liu},
  title     = {A Diffusion Framework for Accurate Fine-Grained Radio Map Reconstruction},
  booktitle = {IEEE Global Communications Conference},
  year      = {2025},
  doi       = {10.1109/GLOBECOM59602.2025.11432702},
}

@inproceedings{rohmtgp2025,
  author    = {Haoxian Liu and Kai Chen and Zezhong Zhang and Li Wang and Shuguang Cui},
  title     = {Land Feature Aware Radio Environment Map Construction Using Radio Oriented Heterogeneous Multitask Gaussian Process},
  booktitle = {IEEE Global Communications Conference},
  year      = {2025},
  doi       = {10.1109/GLOBECOM59602.2025.11432161},
}

@inproceedings{sslradio2025,
  author    = {Jianping Ma and Zezhong Zhang and Junting Chen and Kai Chen and Shuguang Cui},
  title     = {Self-Supervised Learning Informed Radio Environment Map Estimation with Few Samples},
  booktitle = {IEEE Global Communications Conference },
  year      = {2025},
  doi       = {10.1109/GLOBECOM59602.2025.11431663},
}

\end{document}